\documentclass[a4paper]{jpconf}
\usepackage{graphicx}
\usepackage{amsmath,amssymb}
\usepackage{bm}
\usepackage{subfigure}
\usepackage{verbatim}
\usepackage{wrapfig}
\usepackage[dvips]{color} 
\def\slashchar#1{\setbox0=\hbox{$#1$} % set a box for #1
\dimen0=\wd0 % and get its size
\setbox1=\hbox{/} \dimen1=\wd1 % get size of /
\ifdim\dimen0>\dimen1 % #1 is bigger
\rlap{\hbox to \dimen0{\hfil/\hfil}} % so center / in box
#1 % and print #1
\else % / is bigger
\rlap{\hbox to \dimen1{\hfil$#1$\hfil}} % so center #1
/ % and print /
\fi}

\begin{document}
\title{Quark-hadron phase transition in a three flavor PNJL model}

\author{Kanako Yamazaki$^1$, T. Matsui}

\address{Institute of Physics, University of Tokyo, Komaba, Tokyo, 153-8902, Japan}

\ead{$^1$ kyamazaki@nt1.c.u-tokyo.ac.jp}

\begin{abstract}

We study the quark-hadron phase transition by using a three flavor Nambu-Jona-Lasinio model with the Polyakov loop at zero chemical potential, extending our previous work with two flavor model. We show that the equation of state at low temperatures is dominated by pions and kaons as collective modes of quarks and anti-quarks.  As temperature increases, mesonic collective modes melt into the continuum of quark and anti-quark so that hadronic phase changes continuously to the quark phase where quark excitations dominate pressure.
\end{abstract}

\section{Introduction}

Studying the quark-hadron phase transition is one of the most fundamental problem in modern nuclear physics.  
Although many works have been done on the QCD phase transitions by effective models and lattice calculations, there are still many uncertainties especially in the intermediate region between hadronic phase and quark phase. 
The goal of this work is to study the behavior of the equation of state in order to investigate how the degrees of freedom 
change from those of hadrons to quarks and gluons with increasing temperature.  
We work with an effective model written in quark and anti-quark fields and take into account correlations between quarks and anti-quarks to describe mesons. 
Since we wish to describe the chiral and de-confining transitions simultaneously, we use the Nambu-Jona-Lasinio model with the Polyakov loop (PNJL model)\cite{Fukushima:2003fw}. 
In this report, we present a brief account of the results of calculations,
extending our previous work with the two flavor model\cite{Yamazaki:2012ux}, with strangeness degree of freedom. 
It is shown that the equation of state is transformed continuously from that of a gas of light hadrons (pions and kaons) at low temperatures, to a gas of light quarks (u-, d- and s-quarks) and their anti-particles with gluon excitation at high enough temperatures. 
Full account of this work will be reported elsewhere\cite{Yamazaki2013}.

\section{Model setting}

%In this section, we set a three flavor PNJL model and lead an effective action by the path integral method. 
We start with the Lagrangian of a three flavor NJL model in external temporal color gauge fields:  
\begin{eqnarray}
\mathcal{L}=\sum_{i,j=1}^3\bar{q}_i(i\slashchar{D}-\hat{m})_{ij}q_j+\mathcal{L}_4+\mathcal{L}_6
\end{eqnarray}
where
\begin{eqnarray}
\mathcal{L}_4=G\sum_{a=0}^8\bigl[ (\bar{q}\lambda^aq)^2+(\bar{q}i\gamma_5\lambda^aq)^2\bigr] 
\end{eqnarray}
and
\begin{eqnarray}
\mathcal{L}_6&=&-K\bigl[ \mbox{det}\ \bar{q}(1+\gamma_5)q+\mbox{det}\ \bar{q}(1-\gamma_5)q\bigr] \label{L6}
\end{eqnarray}
for three flavor light quarks, $\bar{q}=(\bar{q}_1, \bar{q}_2, \bar{q}_3)=(\bar{u}, \bar{d}, \bar{s})$.  $D_{\mu }=\partial _{\mu }+gA_0\delta _{\mu , 0}$ where $A_0$ is the temporal component of gauge fields, 
$A_0=-iA_4$. 
The gauge field is not treated as a dynamical variable; it plays a role of an external parameter like imaginary color-dependent chemical potential for quarks, and we will take a statistical average over $A_4$ later to define the PNJL model.  
 $\hat{m}$ is a 3 by 3 mass matrix, giving bare masses of u, d and s quarks. 
The partition function is given by
\begin{eqnarray}
Z(T, A_4)=\int [dq][d\bar{q} ]\mbox{exp}\bigl[ \int_0^{\beta }d\tau \int d^3x \mathcal{L} (q, \bar{q}, A_4)\bigr]  . \label{PF}
\end{eqnarray}
With the interaction terms, we build pseudo scalar mesons and scalar mesons from this model. 
For pseudo scalar mesons, there are nine mesons, three kinds of $\pi$, four kinds of $K$, $\eta$ and $\eta'$. 
They make a nonet in SU(3) flavor classification.
In the chiral limit,  all mesons becomes the massless Nambu-Goldstone modes. 
The 6-point interaction, Eq.(\ref{L6}), breaks the axial U(1) symmetry and makes $\eta^0$ massive $\eta '$ meson. 
In addition, the SU(3) flavor symmetry breaking by non-vanishing bare quark masses reproduces the physical mass splitting of 
strange and non-strange mesons.  
There appear also nine scalar mesons, $\sigma $, four kinds of $\kappa $, three kinds of $a_0$, and $f_0$ \cite{Ishida:1997wn, Fariborz:2009cq}. 
They also form a nonet, although not all of them have been confirmed by experiments.

The three flavor NJL type models have not only 4-point interaction of fermion fields but also 6-point interaction. 
%So as to perform the fermion integral of the partition function Eq.(\ref{PF}), 
We replace the 4 and 6 point interactions by Yukawa couplings to bosonic auxiliary fields, extending the method of Hubbard and Stratonovich \cite{Hu59, St57} to 6-point interaction, and perform integral over the quarks field, converting the original Lagrangian of quark fields to that of scalar mesons $\phi_a$ and pseudo scalar mesons $\pi _a$.

The simplest way to calculate the equation of state by the PNJL model is to perform the mean field approximation. 
However in this approximation, meson fields are treated as uniform back ground fields, in other words, thermal fluctuations of mesons are not included under the mean field approximation even at low temperatures.  
We therefore must go beyond the mean field approximation, taking mesonic correlation into account to obtain the equation of state of a meson gas.

In order to calculate a pressure of mesonic correlation, we expand an effective action up to the second order of fluctuations around a stationary point given by the non-vanishing uniform $\phi_0$ field.
All other components of the scalar fields and pseudo scalar fields are set to zero at the stationary point.  
\begin{eqnarray}
I(\phi , \pi , A_4)=I_0+\frac{1}{2}\left. \frac{\delta ^2 I}{\delta \phi _a\delta \phi _b}\right|_{\phi =\phi _0} \phi_a \phi_b 
\ +\left. \frac{1}{2}\frac{\delta ^2 I}{\delta \pi _a\delta \pi _b}\right|_{\phi =\phi _0} \pi_a \pi_b \ \cdots 
\label{eff_act}
\end{eqnarray} 
and the thermodynamic potential is written by  
\begin{eqnarray}
\Omega (T , A_4)=T\Bigl( I_0+\frac{1}{2}\mbox{Tr}_M\mbox{ln}\frac{\delta ^2I}{\delta \phi_a\delta \phi_b} +\frac{1}{2}\mbox{Tr}_M\mbox{ln}\frac{\delta ^2I}{\delta \pi_a\delta \pi_b}\Bigr) . \label{TP}
\end{eqnarray}

\section{Equation of state}

\subsection{Mean field approximation}

Under the mean field approximation, pressure is given by 
\begin{eqnarray}
p_{MF}(T)=p_{MF}^0+2\times 3\sum_f \frac{d^3p}{(2\pi )^3} \frac{\mathbf{p}^2}{3E_f}f_{\Phi }(E_f)
-\mathcal{U}(T, \Phi )
\end{eqnarray}
\begin{wrapfigure}{r}{70mm}
\begin{center}
\includegraphics[clip,width=60mm, angle=270]{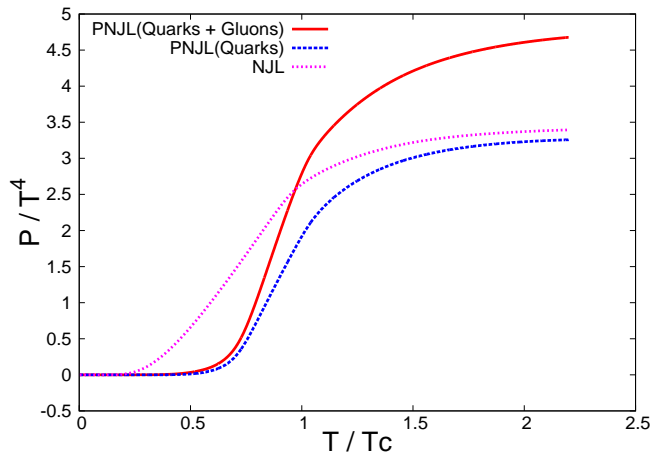}
\end{center}
\caption{
Pressures under the mean field approximation. 
}
\label{fig:PMF}
\end{wrapfigure}
where the first term is a vacuum pressure that doesn't depend on temperature explicitly, 
the second term depends on the temperature through the quark distribution function $f_{\Phi}(E_f)$. 
The third term is an effective potential of gluon, given by hand in order to give the pressure of gluon at high temperatures. The effect of the Polyakov loop appears in the second and third terms. In the quark distribution function, the Polyakov loop works to suppress quark excitations at low temperatures. 

We show the pressure scaled by $T^4$ as a function of temperature in Fig.(\ref{fig:PMF}). The red and the blue lines are calculated by the PNJL model: the red line contains the contribution from an effective potential $\mathcal{U}(T, \Phi)$ while the blue one doesn't so that all contributions of the blue line at high temperatures come only from quarks. 
The pink line is calculated by the NJL model. 
Since the NJL model has no mechanism for the quark confinement, quark excitations are not forbidden even in the hadronic phase at low temperatures. 
In the PNJL model, there is a phase cancellation of the distribution functions of three different color quark states at low temperature.  This cancellation is however not complete and the color neutral excitations with three colored quarks are still allowed. 
It looks like a baryonic excitation but the number of degrees of freedom is not what is expected for baryons\cite{Yamazaki:2012ux}.

\subsection{Mesonic correlation}

\begin{figure}[htbp]
\begin{center}
\includegraphics[clip,width=20mm]{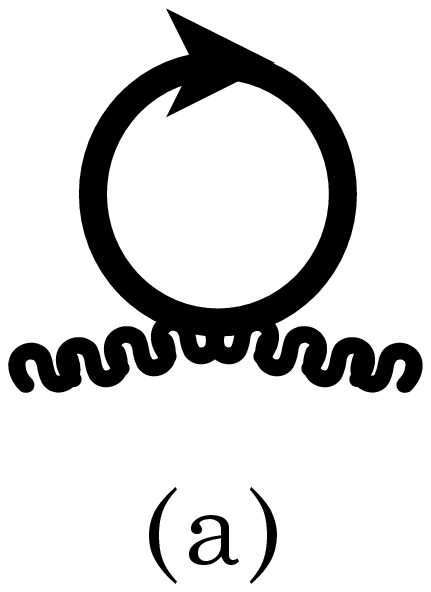}
\includegraphics[clip,width=30mm]{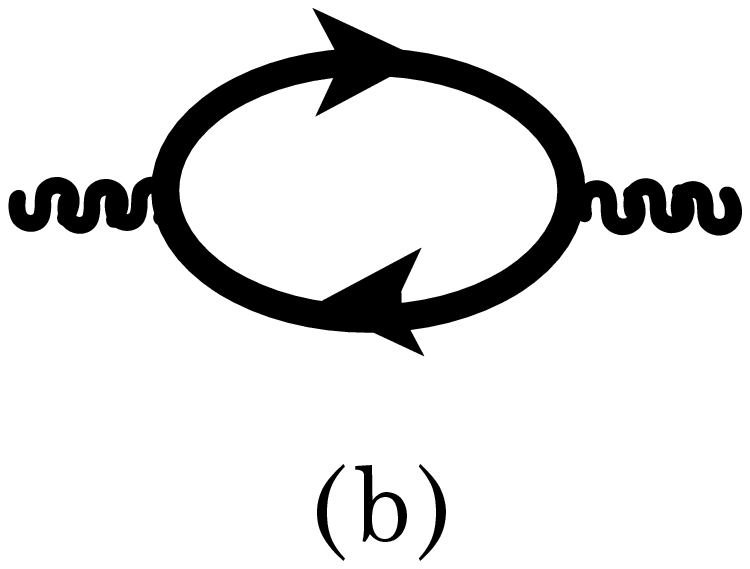}
\end{center}
\caption{
Meson self energy, $\Pi _M$.
}
\label{fig:2point}
\end{figure}

In order to obtain the pressure from mesonic correlations, we calculate the second and the third terms of Eq.(\ref{TP}).  
From the relation between the thermodynamic potential and the pressure, $p=-\Omega /V $,  the pressure of mesonic correlations is given by 
\begin{eqnarray}
p_{M}=-\sum _n \int \frac{d^3q}{(2\pi )^3} \Bigl\{ 3\mbox{ln}\mathcal{M}_{\pi }(\omega _n, q)+4\mbox{ln}\mathcal{M}_{K}(\omega _n, q)+\mbox{ln}\mathcal{M}_{\eta }(\omega _n, q)+\mbox{ln}\mathcal{M}_{\eta ' }(\omega _n, q)\nonumber \ \ \ \ \ \ \ \ \\
+\mbox{ln}\mathcal{M}_{\sigma }(\omega _n, q)+4\mbox{ln}\mathcal{M}_{\kappa }(\omega _n, q)+3\mbox{ln}\mathcal{M}_{a_0}(\omega _n, q)+\mbox{ln}\mathcal{M}_{f_0 }(\omega _n, q)\Bigr\} \label{p_mesons}\ \ \ \ \ \ \ \ 
\end{eqnarray}
where
\begin{eqnarray}
\mathcal{M}_M(\omega_n, q)=\frac{1}{2K'_M}-\Pi _M(\omega_n, q). \label{M} 
\end{eqnarray}
%and
%\begin{eqnarray}
%\Pi_M(\omega_n, q)=\Pi_M^1(A_4)+\Pi_M^2(\omega_n,q, A_4). \label{Pi}
%\end{eqnarray}
%
with $K'_M$ the effective coupling consisting of the four point coupling $G$ and the six point coupling $K$ with appropriate weights for each meson. 
$\Pi _M$ may be decomposed into two terms as indicated in Fig.\ref{fig:2point}.
In Eq.(\ref{p_mesons}), the first four terms are contributions of pseudo scalar mesons and the last four terms of scalar mesons. The pre-factors of $\pi $, $K$, $\kappa $ and $a_0$ are degeneracies of these mesons. 

\begin{wrapfigure}{r}{70mm}
\begin{center}
\includegraphics[clip,width=60mm, angle=270]{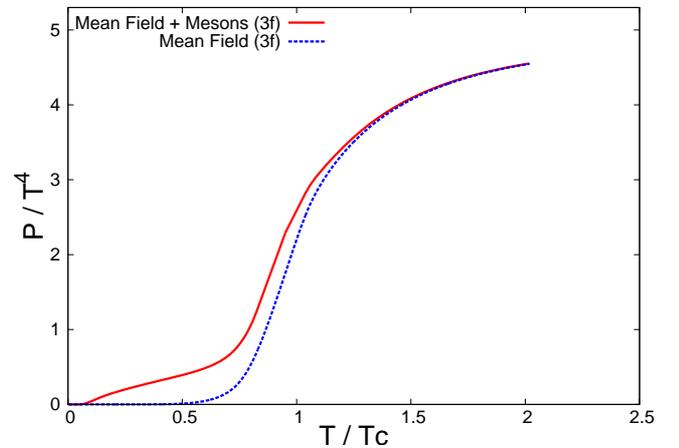}
\end{center}
\caption{
Pressure scaled by $T^4$ as a function of temperature. Red line: pressure from both contributions from the mean field and mesonic correlations. Blue line: pressure without mesonic correlations. The pseudo critical temperature $T_c$ is decided in the mean field level. 
}
\label{fig:EOS_3f}
\end{wrapfigure}
The behavior of the pressure is shown in Fig.(\ref{fig:EOS_3f}). 
At low temperatures,  the pressure is dominated mesonic correlations especially pions and kaons. 
As temperature increases, it approaches quark mean field pressure and corresponds to it at high temperatures. It means that mesons as collective modes at low temperatures melt as temperature increases and resolve to quarks finally.

\section{Conclusion}

We study the quark-hadron phase transition by using a three flavor PNJL model which contains the order parameters of both the chiral phase transition and the de-confining phase transition.  
Under the flavor SU(3) symmetry breaking, four kinds of pseudo scalar mesons($\pi $, $K$, $\eta $ and $\eta '$) and scalar mesons($\sigma $, $\kappa $, $a_0$, and $f_0$) appear with different masses and the equations of state are dominated by mesonic correlations, especially pions and kaons, at low temperatures. 
As temperature increases, the contribution from mesonic correlations decreases and the equation of state is dominated by quarks and antiquark excitations, with gluon excitations added by hand. 
Mesons melt into the continuum of quark-antiquark excitations and the degrees of freedom change from those of hadrons to quarks and gluons.

\section*{Acknowledgements}
We thank members of Komaba Nuclear Theory Group for their interests in this work.  
KY's work has been supported by the University of Tokyo Grants for Ph.D Research, 
Research Assistant for Creation of the Research Core in Physics,  
and School of Science Grants for PhD Students.   
TM's work has been supported by the Grant-in-Aid \# 25400247 of MEXT, Japan.

\end{document}